\documentclass{bioinfo}
\copyrightyear{2022}
\pubyear{2022}

\usepackage{graphicx}
\usepackage{hyperref}
\usepackage{multirow}
\usepackage{url}
\usepackage{tabularx}
\usepackage{amsmath}
\usepackage{amssymb}
\usepackage[ruled,vlined]{algorithm2e}

\SetCommentSty{mycommfont}
\SetKwComment{Comment}{$\triangleright$\ }{}

\usepackage{natbib}
\DeclareMathOperator*{\argmax}{argmax}

\begin{document}
\firstpage{1}

\title[Satellite repeat finder]{De novo reconstruction of satellite repeat units from sequence data}
\author[Li]{Yujie Zhang$^1$, Justin Chu$^{2,3}$, Haoyu Cheng$^{2,3}$, Heng Li$^{2,3}$}
\address{$^1$Harvard School of Public Health, 677 Huntington Avenue, Boston, MA 02115, USA,
$^2$Department of Data Science, Dana-Farber Cancer Institute, 450 Brookline Ave, Boston, MA 02215, USA,
$^3$Department of Biomedical Informatics, Harvard Medical School, 10 Shattuck St, Boston, MA 02115, USA}

\maketitle

\begin{abstract}
Satellite DNA are long tandemly repeating sequences in a genome and may be organized as
high-order repeats (HORs). They are enriched in centromeres and are
challenging to assemble. Existing algorithms for identifying satellite repeats
either require the complete assembly of satellites or only work for simple
repeat structures without HORs. Here we describe Satellite Repeat Finder (SRF), a new
algorithm for reconstructing satellite repeat units and HORs from
accurate reads or assemblies without prior knowledge on repeat structures.
Applying SRF to real sequence data, we showed that SRF could reconstruct known
satellites in human and well-studied model organisms. We also found satellite
repeats are pervasive in various other species, accounting for up to 12\% of
their genome contents but are often underrepresented in assemblies. With the
rapid progress on genome sequencing, SRF will help the annotation of new
genomes and the study of satellite DNA evolution even if such repeats are not
fully assembled.
\end{abstract}

\section{Introduction}

Satellite DNA (SatDNA) are long tandemly repeating sequences that look like
``{\sf BBBBBB$\cdots$}'', where each symbol ``{\sf B}'' represents a
repeat unit, also known as a \emph{monomer}. A monomer ``{\sf B}'' could range
from a few basepairs (bp) to thousands of bp in length and an entire SatDNA could span
megabases in large genomes. Several percent of the human genome, or a couple of
hundred megabases in total, is composed of SatDNA~\citep{Altemose:2022tv}.
Monomers in a SatDNA array are similar in sequence but often not identical due
to random mutations.

In some species, SatDNA may be organized as high-order repeats
(HORs; \citealt{Miga:2019aa}).  For example, the centromere of human chromosome 2
has a pattern like ``{\sf ABCDABCDABCD$\cdots$}''. Letters ``{\sf A}''--``{\sf
D}'' correspond to four diverged alpha repeat monomers of $\sim$171bp each,
respectively, and the ``{\sf ABCD}'' unit is repeated many times in the
centromere with all copies being similar to each other. Researchers who study
centromere repeats usually say ``{\sf ABCD}'' is a 4-mer HOR unit. Because we
will often mention short nucleotide sequences in this article, we will call
``{\sf ABCD}'' in this example as 4-monomer HOR unit to avoid confusion.

SatDNA is often not assembled in long contigs due to its repetitiveness.
We would have to reconstruct SatDNA from raw sequence reads in this
case~\citep{Lower:2018aa}.  Wei et al~\citep{Wei:2014vl} developed k-Seek to
study SatDNA consisting of 2--10bp repeat units. Melters et
al~\citep{Melters:2013va} applied Tandem Repeat Finder (TRF;
\citealt{Benson:1999aa}) to Sanger reads and fragmented short-read contigs to
find the most common monomer in each species.  TAREAN~\citep{Novak:2017wx} does
all-vs-all comparison between short reads, clusters the reads and then
identifies circular structures from the cluster graphs. These methods can
reconstruct unknown monomers but they are unable to reveal HOR structures. On
the contrary, Alpha-CENTAURI~\citep{Sevim:2016tl} reconstructs HORs from long
reads but it requires known monomer sequences.

With improved sequencing technologies, it is now possible to assemble through
human centromeres~\citep{Nurk:2022up}. More recent methods, including
NTRprism~\citep{Altemose:2022tv}, HORmon~\citep{Kunyavskaya:2022tx} and
HiCAT~\citep{Gao2022.08.07.502881}, can identify detailed
chromosome-specific HOR patterns from complete SatDNA sequences in human. These
methods demand high-quality assembly and rich prior knowledge on SatDNA in the
studied species. However, the finished human genome, CHM13, was derived from a
near homozygous molar cell line that is easier to assemble. For a normal
diploid human individual, we could only assemble through a fraction of SatDNA
even with the best possible data and
algorithm~\citep{Rautiainen:2023aa}. The complete assembly of other
species is even rarer. This has limited the application of such assembly-based
SatDNA reconstruction algorithms.

In this article, we will describe a new algorithm, Satellite Repeat Finder
(SRF), for assembling SatDNA repeat units. SRF overcomes the limitation of
previous methods. It is applicable to both accurate reads and high-quality
assembly and is able to automatically reconstruct HORs with no prior knowledge
on monomer sequences.

\section{Results}

\subsection{The SRF algorithm}

In a SatDNA array ``{\sf BBBBBB$\cdots$}'', suppose every monomer ``{\sf B}''
is identical to each other. Under a long enough $k$, the $k$-mer de Bruijn
graph of the SatDNA array will be a single cycle.  When there are
basepair differences between monomers, the de Bruijn graph will not be a simple cycle.
If there are many different monomers, the de Bruijn graph can become very
complex and cannot be resolved with classical graph cleaning
algorithms~\citep{Zerbino:2008uq}.

Our intuition is that if there are many copies of the monomer, we may still be
able to find a cycle composed of highly abundant $k$-mers in the de Bruijn
graph. We can start with the most abundant $k$-mer and at each bifurcation in
the graph, we greedily choose the $k$-mer of the highest occurrence. We repeat
this process until we go back the starting $k$-mer, which will reconstruct a
repeat unit, or come to a deadend, which will be discarded.
Algorithm~\ref{alg:srf} provides more details. Here, $t\to s$ indicates $k$-mer
$t$ and $s$ are adjacent in the de Bruijn graph. For simplicity, this algorithm
traverses a unidirected de Bruijn graph. In SRF, we implemented a bidirected
de Bruijn graph such that we will not find a repeat unit on both strands.

SRF works with Illumina short reads, PacBio HiFi long reads and high-quality
assembly contigs and can identify HORs (Table~\ref{tab:tool}).  When assembling
satellite repeats from PacBio HiFi reads in this article, we counted 151-mers
with KMC~\citep{Kokot:2017aa} and collected 151-mers occurring $\ge$10 times
over the average read coverage. K-mer counting may take a few tens of minutes
for a high-coverage human dataset and is the performance bottleneck. SRF only
takes seconds to reconstruct all repeat units after k-mer counting.

\begin{table}[!hb]
\processtable{Features of user-facing tools for SatDNA reconstruction}
{\label{tab:tool}
\begin{tabular}{p{3.6cm}cccc}
\toprule
Tool & Reads & Contigs & De novo & HORs \\
\midrule
Alpha-CENTAURI & Yes & No & No & Yes \\
HiCAT    & No & Yes & No & Yes \\
HORmon   & No & Yes & No & Yes \\
NTRprism & No & Yes & Yes  & Yes \\
SRF (this work) & Yes & Yes & Yes  & Yes \\
TAREAN   & Yes& No  & Yes  & No \\
\botrule
\end{tabular}}{``Reads'': whether the tool works with unassembled reads.
``Contigs'': whether the tool works with high-quality contigs. ``De novo'':
whether the tool works without known monomer sequence as input. ``HORs'':
whether the tool can identify high-order repeats (HORs). References to these
tools can be found in the Introduction section.}
\end{table}

\subsection{Estimating satellite abundance}

The SRF algorithm does not provide a good estimate of repeat abundance. We
mapped all input sequences against reconstructed repeat units to measure the
total length of each repeat unit.
For human CHM13 data, we observed many diverged hits
between HORs and scattered monomers in pericentromeric regions. We thus filtered
hits of identity below 90\% to get more accurate HOR length estimates.
The effect of the identity is determined by the repeat structure in a species.
For example, switching off the filter would increase the total abundance
estimate by 40\% for human but only by 4\% for \emph{A.thaliana}. We still
applied this filter to all datasets even though this may lead to underestimates
for some species.

SRF may reconstruct repeat units similar in sequence. The similar repeat units
may be mapped the same genomic locus. To remove redundancy, we only select the
hit of the highest identity among hits overlapping on an input sequence.
With this procedure, we map each base on an input sequence to at most one
repeat unit.

Occasionally a small number of long terminal repeats (LTRs) may occur tandemly
in a few region. SRF may identify such LTRs even though they do not form long
tandem arrays. When estimating abundance, we additionally filter out repeat
with $<$2 tandem copies in the middle of a sequence or with $<$1.5 tandem
copies when the repeat-to-read alignment reaches the end of a read. This filter
is reliable when we apply SRF to assemblies but may miss long repeat units when
applied to reads. We again opted for conservative estimates.

\subsection{A brief introduction to human satellites}

In the human genome, the most abundant satellite family is alpha satellites
with most of them present in long alpha HORs ($\alpha$HORs). The
consensus of the minimal alpha repeat unit is 171bp in length. The active
centromeric regions that centromeric proteins bind to are primarily
composed of $\alpha$HORs~\citep{Altemose:2022tv}. Conversely, though, not all
$\alpha$HORs are present in the active regions. These inactive $\alpha$HORs
tend to be shorter than active ones. Alpha repeat monomers are also present in
pericentrometic regions without clear HOR structures. In addition to alpha
repeat, the human genome is also enriched with three types of human satellites
(HSat1--3), contributing to a few percent of human
genome~\citep{Altemose:2022vw}. Almost all these satellites are located around
centromeres or on the long arm of the Y chromosome.

In the human reference genome GRCh38~\citep{Schneider:2017aa}, $\alpha$HORs
were computationally generated from a Hidden Markov Model~\citep{Miga:2014aa};
HSats are underrepresented. At present, only the T2T-CHM13
assembly~\citep{Nurk:2022up} provides a complete representation of all
satellite arrays.

\begin{algorithm}[tb]
\DontPrintSemicolon
\footnotesize
\KwIn{$k$-mer set $\mathcal{S}$ and count function $\phi:\mathcal{S}\to\mathbb{N}$}
\KwOut{SatDNA repeat units}
\BlankLine
\While {$\mathcal{S}\not=\emptyset$} {
	$t_0\gets\argmax_{s\in\mathcal{S}}\phi(s)$\Comment*[r]{Most abundant $k$-mer in $\mathcal{S}$}
	$\mathcal{H}\gets\{t_0\}$\;
	\While {$\mathcal{H}\not=\emptyset$} {
		$t\gets\argmax_{s\in\mathcal{H}}\phi(s)$\Comment*[r]{Most abundant $k$-mer in $\mathcal{H}$}
		$\mathcal{H}\gets\mathcal{H}\setminus\{t\}$\Comment*[r]{Remove $t$ from $\mathcal{H}$}
		\If (\Comment*[f]{Back to the starting $k$-mer}) {$t=t_0$ {\bf and} $t_0\not\in\mathcal{S}$} {
			Trace back from $t$ and report a repeat unit\;
			{\bf break}\;
		}
		\For (\Comment*[f]{Traverse $t$'s neighbors}) {$s\in\mathcal{S}$ {\bf and} $t\to s$} {
			$\mathcal{S}\gets\mathcal{S}\setminus\{s\}$\;
			$\mathcal{H}\gets\mathcal{H}\cup\{s\}$\Comment*[r]{Move $s$ from $\mathcal{S}$ to $\mathcal{H}$}
			${\rm parent}(s)=t$\Comment*[r]{Keep parent for traceback}
		}
	}
	$\mathcal{S}\gets\mathcal{S}\setminus\{t_0\}$\Comment*[r]{Remove $t_0$ from $\mathcal{S}$}
}
\caption{Assemble SatDNA repeat units}\label{alg:srf}
\end{algorithm}

\begin{table*}[!tb]
\processtable{Human chromosome-specific high-order alpha repeats ($\alpha$HORs)}
{\label{tab:eval}
\begin{tabular*}{\textwidth}{@{\extracolsep{\fill}}rrrlllll}
\toprule
\multirow{2}{*}{chr} & \multicolumn{1}{l}{HORmon $^a$}& \multicolumn{1}{l}{HiCAT $^b$} & SRF ($k$=171)  & \multicolumn{1}{l}{SRF/171}  & \multicolumn{1}{l}{SRF/171}    & \multicolumn{1}{l}{SRF/101}  & SRF ($k$=171) \\
                     & \multicolumn{1}{l}{centromere} & \multicolumn{1}{l}{centromere} & chromosome $^c$& \multicolumn{1}{l}{assembly $^d$} & \multicolumn{1}{l}{HiFi reads $^d$} & \multicolumn{1}{l}{Illumina $^d$} & HPRC assembly $^e$\\
\midrule
1  & 6  & 2  & 6 (4.2); 11 (0.5) & 6 (2.0)  & 6 (3.5)  & 2 (2.2)  & 6 [89] \\
2  & 4  & 4  & 4 (2.3)           & 4 (2.3)  & 4 (2.2)  & 4 (2.2)  & 4 [94] \\
3  & 17 & 17 & 17 (1.4)          & 17 (1.4) & 17 (1.4) & 17 (1.4) & 17 [94] \\
4  & 19 & 19 & 19 (3.5)          & 19 (3.5) & 19 (2.9) & 19 (3.4) & 19 [94] \\
5  & 6  & 12 & 8 (2.5)           & 8 (1.8)  & 4 (1.9)  & missing  & 8 [43]; 4 [37] \\
6  & 18 & 18 & 18 (2.0)          & 18 (2.0) & 18 (2.0) & 18 (2.0) & 18 [93] \\
7  & 6  & 6  & 6 (3.3)           & 6 (3.2)  & 6 (3.2)  & 6 (3.2)  & 6 [92]; 12 [2] \\
8  & 11 & 15 & 7 (1.1)           & 7 (1.1)  & 7 (1.0)  & 11 (1.0) & 7 [61]; 8 [33] \\
9  & 7  & 11 & 4 (1.8)           & 4 (1.4)  & 11 (2.0) & 4 (1.7)  & 4 [77]; 11 [17] \\
10 & 8  & 6  & 8 (2.1)           & 8 (2.1)  & 8 (1.7)  & 8 (2.1)  & 6 [66]; 8 [28] \\
11 & 5  & 5  & 5 (3.4)           & 5 (3.3)  & 5 (3.4)  & 5 (3.4)  & 5 [94] \\
12 & 8  & 8  & 8 (2.6)           & 8 (2.6)  & 8 (2.6)  & 8 (2.6)  & 8 [94] \\
13 & 11 & 7  & 4 (0.4)           & 4 (0.4)  & 7 (1.5)  & 7 (1.5)  & 4 [55]; 11 [23]; 7 [16] \\
14 & 8  & 8  & 8 (2.6)           & missing  & missing  & missing  & missing \\
15 & 11 & 15 & 11 (0.8); 20 (0.5)& 11 (0.8) & 11 (0.8) & 11 (0.8) & 11 [94] \\
16 & 10 & 10 & 10 (2.0)          & 10 (1.9) & 10 (1.9) & missing  & 10 [94] \\
17 & 16 & 14 & 16 (3.3)          & 16 (3.3) & 16 (3.5) & 16 (3.5) & 16 [56]; 13 [38] \\
18 & 12 & 12 & 8 (3.6)           & 8 (3.8)  & 12 (4.9) & missing  & 12 [66]; 8 [19] \\
19 & 2  & 2  & 4 (0.4); 2 (0.4)  & missing  & 13 (0.5) & missing  & 13 [29]; 32 [4] \\
20 & 16 & 16 & 16 (2.1)          & 16 (2.1) & 16 (2.1) & 8 (0.5)  & 16 [94] \\
21 & 11 & 11 & 11 (0.3)          & missing  & missing  & missing  & missing \\
22 & 8  & 8  & 8 (2.9); 20 (0.5) & 8 (2.8)  & 8 (2.6)  & 8 (2.9)  & 8 [94] \\
X  & 12 & 12 & 12 (3.1)          & 12 (3.1) & 12 (3.1) & 12 (3.1) & 12 [76] \\
Y  & 34 &No Y& 34 (0.3)          & 34 (0.3) & No Y     & No Y     & 34 [18] \\
\botrule
\end{tabular*}
}{$^a$ $\alpha$HOR lengths in the monomer unit in the CHM13 v2.0 genome, retrieved
from~\citet{Kunyavskaya:2022tx}.  $^b$ length of ``top 1'' $\alpha$HOR from each
chromosome retrieved from~\citet{Gao2022.08.07.502881}. Both HORmon and HiCAT
were applicable to extracted centromeric sequences only.$^c$ SRF applied to
each CHM13 chromosome separately. In a format ``$m$ ($L$)'', $m$ denotes the
length of an HOR in the monomer unit and $L$ is its span on the CHM13 assembly
in megabases. $^d$ SRF applied to CHM13 assembly, PacBio High-Fidelity (HiFi) reads and
Illumina short reads, respectively. $k$=101 used for Illumina reads. CHM13
reads do not contain chrY. $^e$ SRF applied to 94 phased haploid assemblies
produced by the Human Pangenome Reference Consortium (HPRC).  In a format ``$m$
[$n$]'', $m$ is the monomer length and $n$ is the number of samples with the
HOR according to manual inspection.} 
\end{table*}

\subsection{Annotating satellite repeats in human T2T-CHM13}

We first ran SRF on each T2T-CHM13
chromosome separately and compared the results to existing annotations by
HORmon~\citep{Kunyavskaya:2022tx} and HiCAT~\citep{Gao2022.08.07.502881}.
HORmon reports the same $\alpha$HOR lengths as \citet{Altemose:2022tv}.
In Table~\ref{tab:eval}, column ``SRF ($k$=171) chromosome'' shows the lengths
of HORs identified by running SRF on individual chromosomes.  SRF
reported the same $\alpha$HOR lengths as HORmon except for chromosome 5, 8, 9,
13 and 18. The chromosome 8 of T2T-CHM13 has been well studied by
\citet{Logsdon:2021aa}. Although the 11-monomer is the most abundant, it is
interleaved with 4-, 7- and 8-monomers that are derived from the 11-monomer.
The 7-monomer is the second most abundant array and forms the longest
$\alpha$HOR array in the middle of the centromere. The greedy SRF algorithm
chooses the 7-monomer over the 11-monomer possibly because the 7-monomer
has a more conservative consensus. The SRF-HORmon differences in other
chromosomes may have a similar cause.

Unlike HORmon and HiCAT which require users to provide the monomer
sequence and prepare centromeric sequences, SRF was directly applied to
whole chromosome sequences with no prior knowledge. In addition to active
$\alpha$HORs, SRF identified shorter $\alpha$HOR arrays outside the active
regions. It also found many long non-alpha satellite arrays including a repeat
unit of 1,814 bp on chr15, of 6,112 bp on chr16, of 3,569 bp on chrY and of
2,420 bp on chrY as well. These span over one megabase and have been reported
previously~\citep{Altemose:2022vw}.

SRF further found a satellite array on the long arm of chromosome 1 between
coordinate 227,746,662 and 228,024,151. The repeat unit is 2,240 bp in length,
composed of an AluY repeat, a 5S-RNA and di-nucleotide repeats. This is the
only non-centromeric array in T2T-CHM13 longer than 100kb.

The SRF inference on the whole T2T-CHM13 genome (column ``SRF/171 assembly'' in
Table~\ref{tab:eval}) is close to the inference on individual chromsomes. SRF
missed the $\alpha$HOR array on chr14 and chr21 because chr22 and chr13,
respectively, have very similar arrays which are merged during the whole-genome
inference.

SRF works on sequence reads which HORmon, HiCAT and NTRprism are not
applicable to. On PacBio High-Fidelity (HiFi) reads, SRF reconstructed
$\alpha$HORs similar to the whole-genome reconstruction (column ``SRF/171 HiFi
reads''). It can also identify the majority of $\alpha$HORs from Illumina short
reads (column ``SRF/101 Illumina'' in Table~\ref{tab:eval}), though the use of
shorter 101-mer reduces the sensitivity to some arrays.

From Table~\ref{tab:eval} we can see that some $\alpha$HOR arrays, such as
those on chr3 and chr11, can be consistently reconstructed by various tools on
different types of input data. However, some other arrays, such as those on
chr8 and chr19, are intrinsically harder to reconstruct. These are probably
because monomers in a HOR may be connected in different ways, as is shown by
\citet{Kunyavskaya:2022tx}.

\subsection{Satellite repeats in multiple human assemblies}

We applied SRF independently to each phased human assembly produced by the
Human Pangenome Reference Consortium (HPRC). We identified $\alpha$HORs
with dna-brnn~\citep{Li:2019aa}, aligned them to the T2T-CHM13 genome and
HORmon consensus and manually assigned the $\alpha$HORs to chromosomes based on
the similarity to existing annotations. The last column of Table~\ref{tab:eval}
shows the $\alpha$HOR lengths and their frequencies. We consider two
$\alpha$HORs are different if they have different lengths and the shorter
$\alpha$HOR cannot be aligned into the longer one at $<$2\% sequence
divergence.

There are 47 diploid samples and 94 haploid assemblies. We could find
$\alpha$HORs assigned to individual chromosomes in most cases. We sometimes see
$\alpha$HORs of different lengths assigned to the same chromosome but their
sequence divergences are small. This again could be caused by the different
ways HOR monomers are connected in individual
samples~\citep{Logsdon:2021aa,Kunyavskaya:2022tx}.

We also applied SRF to the pool of all haploid assemblies. This procedure
may miss infrequent satellite arrays but it helps to simplify the study of
shared arrays. In addition to active $\alpha$HORs, we identified supposedly
inactive $\alpha$HORs that are at $>$10\% divergence from existing annotated
$\alpha$HORs. Notably, there is a 20-monomer $\alpha$HOR that is mapped to
the chromosome 15 of T2T-CHM13 and spans several hundred kilobases in most
samples. There are other examples like this. SRF also found long HSat
arrays and non-HOR satellites, including those found in CHM13, which are easy
to identify as they do not have internal structures.

\subsection{Satellite repeats in \emph{Arabidopsis thaliana}}

We obtained three HiFi datasets (Table~\ref{tab:at}) and downsampled them to about
40-fold coverage each. Col-0N and Ey15-2R were sequenced from a pool of
multiple samples and Col-0R from a single sample.

\begin{table}[!hb]
\processtable{\emph{A. thaliana} PacBio HiFi datasets}
{\label{tab:at}
\begin{tabular}{p{1.5cm}p{1.5cm}ll}
\toprule
Sample & Strain & SRA Accession & Source \\
\midrule
Col-0N & Col-0 & ERR6210723 & \citet{Naish:2021aa} \\
Col-0R & Col-0 & ERR8666127 & \citet{Rabanal:2022aa} \\
Ey15-2R & Eyach15-2 & ERR8666125 & \citet{Rabanal:2022aa} \\
\botrule
\end{tabular}}{}
\end{table}

The \emph{A. thaliana} centromeres are composed of CEN180 satellites which are
known to have high-order organizations~\citep{Naish:2021aa}. When we pooled the
two Col-0 datasets and applied SRF, we reconstructed one 1-monomer, two
2-monomers, one 3-monomer and one 6-monomer HORs. However, this result is
unstable. If we ran SRF on each Col-0 dataset separately or on the full
coverage, we would get different HORs. We thus ignored the HOR structures and
focused on the abundance of CEN180 only.

\begin{figure}[!tb]
\includegraphics[width=.49\textwidth]{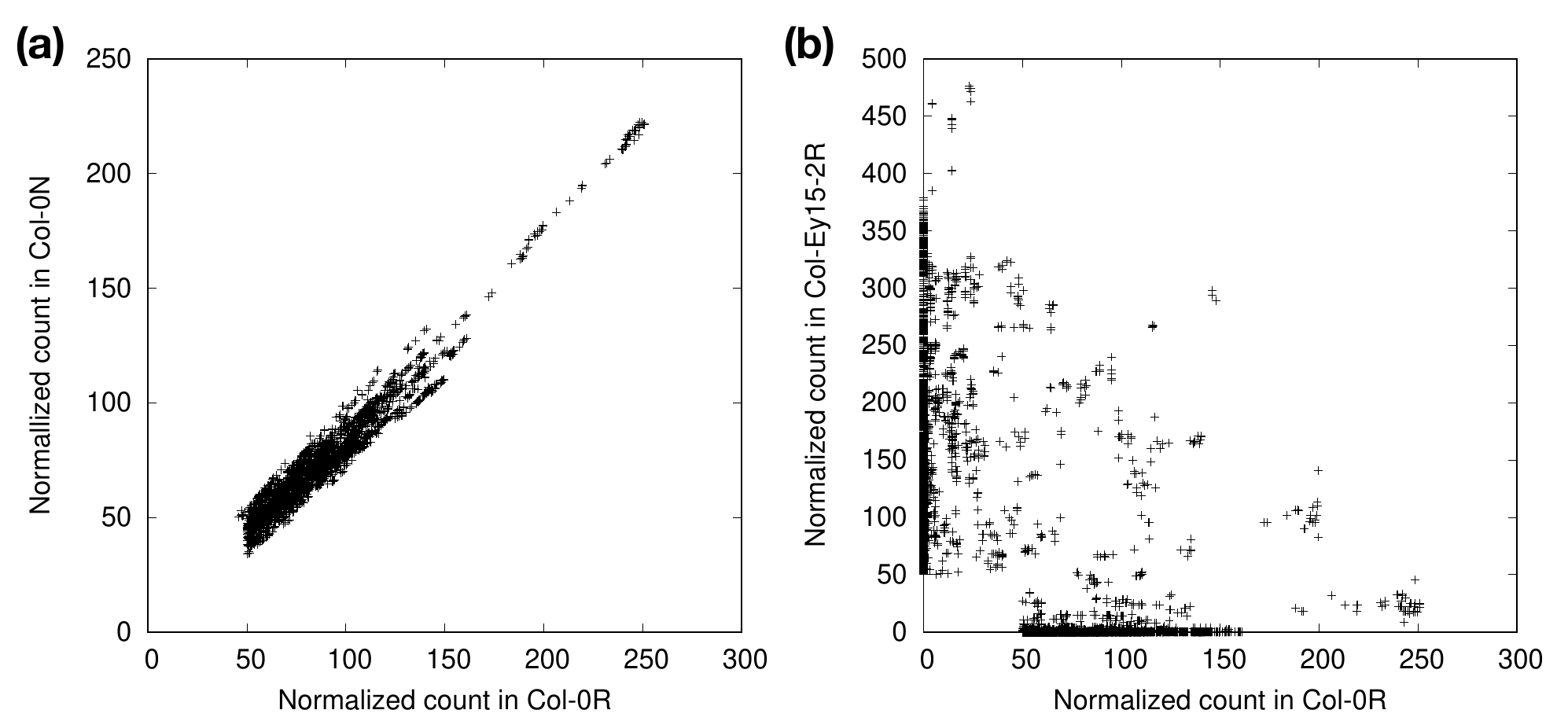}
\caption{Normalized counts of 179-mers in three \emph{A. thaliana} read
datasets. Raw 179-mer counts in reads are normalized by coverage. A 179-mer is
selected in the plot if it matches the CEN180 satellite and if its normalized
count is at least 50 in one of the datasets. {\bf (a)} Counts between two
different samples from the same strain. {\bf (b)} Counts between two different
strains.}\label{fig:1}
\end{figure}

SRF estimated that 5.2\% of read bases in Col-0N and 7.5\% in Col-0R are
composed of CEN180 satellites. Col-0R has 44\% more CEN180 satellites than
Col-0N. To check whether this large difference is caused by an artifact, we
inferred the relative CEN180 abundance with a different approach as follows.
We collected 179-mers matching CEN180 and occurring $\ge$400 times in the
pooled Col-0 dataset. We counted the total numbers of these 179-mers in Col-0N
and Col-0R separately and normalized the counts by read bases. We found Col-0R
has 41\% more CEN180 179-mers than Col-0N, broadly in line with the earlier
analysis. Centromere contents may differ greatly even between two samples from
the same strain.

SRF estimated that 11.5\% of Ey15-2R is composed of CEN180 satellites, higher
than both Col-0R and Col-0N. Furthermore, while Col-0R and Col-0N share similar
high-occurrence 179-mers that match CEN180 (Fig.~\ref{fig:1}a), Col-0 and
Ey15-2 share few common 179-mers (Fig.~\ref{fig:1}b). The centromere sequences
between strains are distinct both in content and in length.


In addition to the CEN180 satellite, SRF also reconstructed a 10,067 bp rDNA
unit from the two Col-0 datasets. It has 3.1\% abundance in Col-0R and 1.7\% in
Col-0N. If we assume the \emph{A. thaliana} genome is 132Mb in length
according to the Col-0N assembly, Col-0R has $\sim$400 copies of this rDNA unit
while Col-0N has $\sim$220 copies. The Col-0N assembly~\citep{Naish:2021aa} only has seven copies,
located towards the telomeric ends of chr2 or chr4 short arms.  SRF did not
reconstruct an rDNA unit from Ey15-2R. We mapped the Col-0 rDNA unit to Ey15-2R
reads and estimated that Ey15-2R has $\sim$200 copies.

\subsection{Satellite repeats in other model organisms}

We applied SRF to the HiFi reads of three model organisms~\citep{Hon:2020aa}: the
reference C57BL/6J strain of \emph{Mus musculus} (mouse; AC:SRR11606870), the
F1 generation of the reference ISO1 strain and the A4 strain of
\emph{Drosophila melanogaster} (AC:SRR10238607), and the B73 strain of \emph{Zea
mays} (maize; AC:SRR11606869).

In mouse, SRF identified two satellite units. The second most abundant repeat
is the 234 bp major satellite around
centromeres~\citep{Arora:2021aa,Thakur:2021aa}. The first is 1,199 bp in length,
composed of 10 copies of the 120 bp minor satellite unit. This confirms the
high-order organization of minor satellites observed by \citet{Pertile:2009aa}.
The full-length hits of this repeat in the mouse reference genome mostly come
from the sex chromosomes and are all below 75\% in identity.  Nonetheless, this
repeat is abundant in reads with the majority of alignments at 95\% identity or
higher. To further investigate this repeat, we assembled the HiFi reads with
hifiasm~\citep{Cheng:2021aa}. We can find long tandem arrays of this repeat on
multiple contigs, all shorter than 1.1Mb.  Hifiasm keeps the repeat content but
is unable to assemble this satellite.

In \emph{Drosophila}, the most abundant satellite SRF identified is a 358 bp
repeat unit hitting 0.90\% of read bases. It belongs to the 1.688
family~\citep{Khost:2017aa}. The abundance of the 358 bp repeat is lower in the
BDGP6 reference genome, at 0.26\% only. SRF assembled the 240 bp Intergenic
spacer (IGS; \citealt{Shatskikh:2020aa}) into two sequences, at 240 bp and
239 bp, respectively. The edit distance between the two IGS sequences is 5. They
hit to 0.43\% of read bases in total but are depleted in the reference at
$<$0.01\% only. SRF also found other known satellite repeats such as (AAGAC)n,
(AACAC)n, (AATAG)n, (GGTCCCGTACT)n and
(AATAACATAG)n~\citep{Shatskikh:2020aa,Thakur:2021aa}.  There are more copies of
these repeat units but because they are short, they contribute less to the
genome in comparison to the 1.688 and IGS satellites.

SRF reconstructed a 5,045 bp repeat unit at 0.34\% abundance in reads and
0.08\% in the reference genome. It harbors histone genes and is located in a
small region on chromosome 2L. To investigate further, we assembled the HiFi
reads using the hifiasm trio-binning mode with ISO1 and A4 short reads from
SRR6702604, SRR457665, SRR457666 and SRR457707. When aligning the ISO1
haplotype assembly to the reference genome, we see a clean 242kb insertion
entirely composed of the 5,045 bp histone repeat. The insertion has 48
tandem copies at $>$99\% identity between the copies. The BDGP6 reference genome might
have misassembled this region.

In maize, SRF reconstructed a 741 bp repeat unit at 0.25\% abundance. It matches
the SAT1\_ZM record in RepBase. This SRF unit includes four copies of a 180 bp
knob-associated repeat~\citep{Ananiev:1998ab}. In the NAM-5.0 reference genome
or the hifiasm assembly, this repeat tends to be present in short contigs and
towards ends of long contigs. It is not assembled well. SRF also identified
many potential repeat units at $<$0.09\% abundance in reads. Nonetheless, none
of them form long tandem arrays. Meanwhile, under the 151-mer setting, SRF failed to identify the 156 bp CentC
repeat~\citep{Ananiev:1998aa}. SRF could find this repeat if we counted
101-mers. Only 0.045\% of read bases were mapped to CentC. Low-abundance SatDNA
is harder to assemble correctly.

\subsection{Comparison to TAREAN}

TAREAN~\citep{Novak:2017wx} can identify novel satellite repeats from sequence
reads. Its developers recommend to use reads at up to 0.5-fold coverage to
avoid redundancy between reads sequenced from the same loci. We
ran TAREAN on simulated short reads at 0.2-fold from the Dropsophila HiFi
dataset described above, without introducing additional sequencing errors.
TAREAN found six high-confidence satellite repeats, including the 1.688 family
and the histone cluster, (GGTCCCGTACT)n and (AATAACATAG)n. The other two TAREAN
repeats also hit to SRF contigs. SRF assembled eight more SatDNA repeat units
at $>$0.05\% abundance. Manually inspecting the alignment of SRF contigs to raw HiFi reads,
we observed tandem pattern for all of them, suggesting they were real SatDNA.

To evaluate whether TAREAN can reconstruct HORs, we ran TAREAN on 0.2-fold
CHM13 reads randomly sampled from SRR2088062. TAREAN took 5 hours and found
four high-confidence satellite repeats, including a 2-monomer alpha repeat at
1.0\% abundance, a HSat2 repeat at 0.9\%, a SAR satellite and a beta satellite.
TAREAN did not identify other HORs.

%
%

\subsection{Satellite repeats in other species}

\begin{table}[bt]
\processtable{HiFi datasets for non-model organisms}
{\label{tab:dtol}
\begin{tabular}{lll}
\toprule
Species & Common name & Source \\
\midrule
\emph{A. bisporus} & cultivated mushroom & PRJEB52214 \\
\emph{A. ruthenus} & sterlet & PRJEB19273 \\
\emph{C. elaphus} & red deer & \citet{pemberton_genome_2021} \\
\emph{C. lupus} & grey wolf & \citet{sinding_genome_2021}\\
\emph{D. lineata} & orange-striped anemone & \citet{wood_genome_2022} \\
\emph{F. $\times$ ananassa} & royal royce strawberry & \citet{Hon:2020aa} \\
\emph{H. helix} & ivy & PRJEB47300 \\
\emph{L. sulphureus} & chicken mushroom & \citet{wright_genome_2022} \\
\emph{M. domestica} & apple & \citet{konyves_genome_2022} \\
\emph{M. meles} & Eurasian badger & \citet{newman_genome_2022} \\
\emph{M. sylvestris} & crab apple & \citet{ruhsam_genome_2022} \\
\emph{M. glacialis} & spiny starfish & \citet{lawniczak_genome_2021} \\
\emph{O. orca} & killer whale & \citet{foote_genome_2022} \\
\emph{O. sativa} & rice & SRR10238608 \\
\emph{P. pellucida} & blue-rayed limpet & \citet{lawniczak_genome_2022} \\
\emph{R. muscosa} & yellow-legged frog & \citet{Hon:2020aa} \\
\emph{V. atalanta} & red admiral butterfly & \citet{lohse_genome_2021} \\
\botrule
\end{tabular}}{}
\end{table}

\begin{figure}
\includegraphics[width=.49\textwidth]{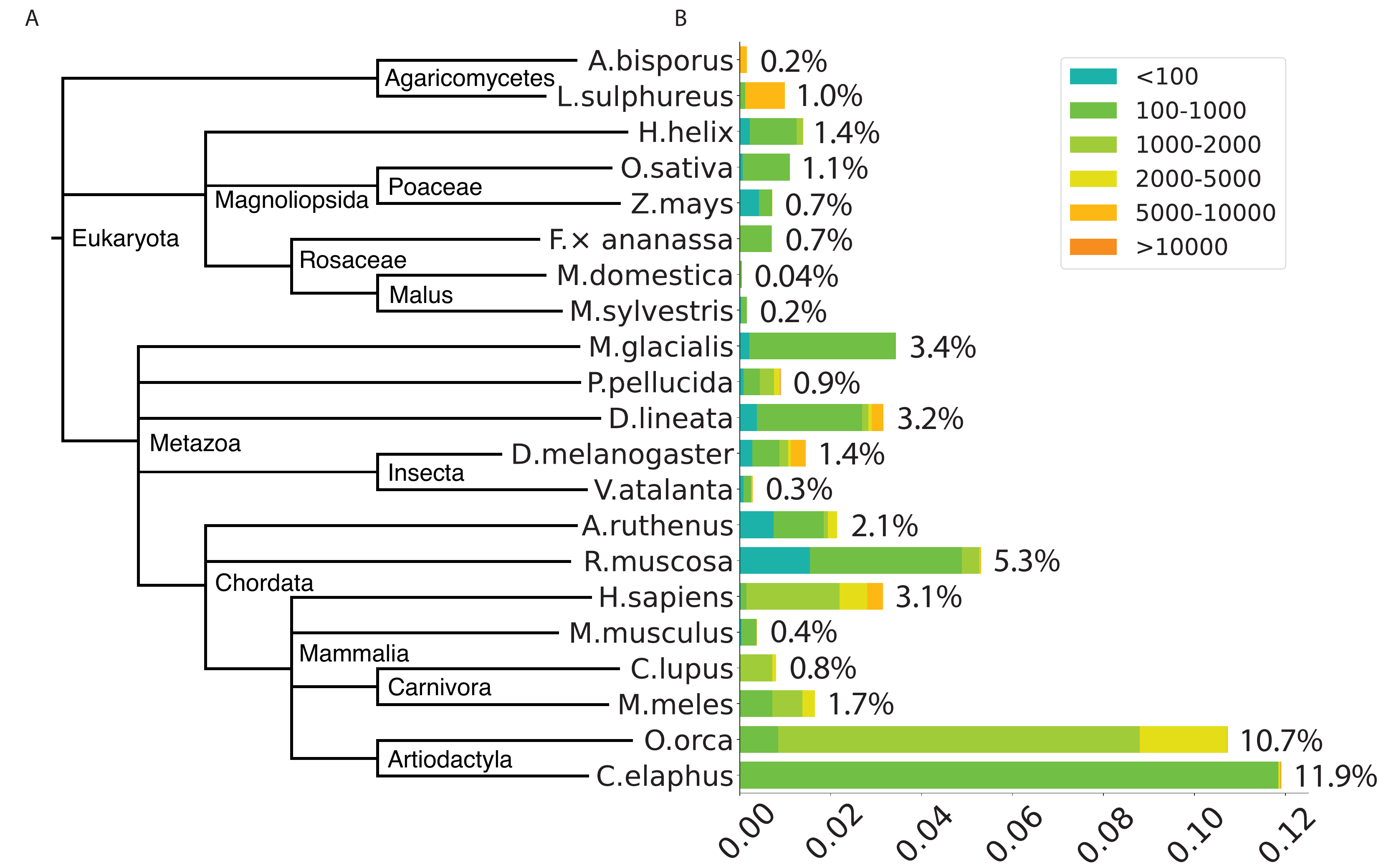}
\caption{Abundance of satellite DNA in 21 species.}\label{fig:dtol}
\end{figure}

We randomly selected 14 species from the Darwin Tree of Life project and
collected two species from \citet{Hon:2020aa} (Table~\ref{tab:dtol}). We assembled
SatDNA in these and several other species described in earlier sections.
SRF may reconstruct mitochondria or chloroplast from sequence reads.
We manually removed them based on NCBI BLAST against the nt database.
We then estimated the abundance of SatDNA in each of these species (Fig.~\ref{fig:dtol}).

Red deer (\emph{C. elaphus}) has the highest abundance at 11.9\%. A single
796 bp repeat unit accounts for 10.3\% of satellite DNA. Killer whale (\emph{O.
orca}) in the same order is also enriched with satellite DNA. Yellow-legged
frog (\emph{R. muscosa}) is next to killer whale. SRF reconstructed many
distant variants of a 131 bp repeat unit. On the other extreme, apples (\emph{M.
domestica} and \emph{M. sylvestris}) barely have satellite repeats partly
because they have transposon-rich centromeres~\citep{Zhang:2019ab}. Plants and
fungi are generally depleted of satellites.

It is worth noting that our abundance estimate may be an underestimate due to
the additional filters we used. For example, chicken mushroom (\emph{L.
sulphureus}) had a repeat unit of 9,659 bp at 0.9\% abundance. As we discarded
alignments shorter than 1.5 times 9,659 bp, we filtered out many HiFi reads
shorter than this threshold even if entire reads were aligned to the repeat.
The abundance estimate would be doubled without this filter. Such long repeat
units are infrequent in the species we studied.

To investigate what satellites are organized as HOR, we ran
TRF~\citep{Benson:1999aa} on SRF-assembled repeat motif. A repeat motif is
considered to have a high-order structure if TRF identifies a tandem repeat
repeating at least three times and covering 90\% of the motif. With this
criterion, 98.5\% of human satellites are HORs with a variety of number of
monomers. 8.6\% of satellites in Eurasian badger (\emph{M. meles}) are HORs of
a 138 bp monomer, contributing to 0.14\% of the genome. The other species in
our survey either do not have multiple HORs composed of similar monomers or
only have HORs at $<$0.1\% abundance. Consistent with our observation,
\citet{Melters:2013va} rarely identified HORs consisting of $\ge$3 monomers.
The authors attributed this to the limited Sanger read length. Based on longer
reads and a different algorithm, our result suggests that most species do not
exhibit rich HOR structures.

\section{Discussions}

SRF is a de novo assembler for reconstructing SatDNA repeat units and can
identify most known HORs and SatDNA in well-studied species without prior
knowledge on monomer sequences or repeat structures. It is the only de novo
algorithm for reconstructing HORs from sequence reads as well as high-quality
assemblies. SRF only depends on a third-party k-mer counter. It is easy to run
and fast to execute.

SRF uses a greedy algorithm to assemble SatDNA repeat units. When two repeat
units share long similar sequences, the one of lower abundance and higher
diversity may be missed. We plan to improve the current algorithm by reporting
multiple overlapping cycles. This may be able to find a more complete
collection of HORs in the human genome.

Meanwhile, although SRF can reconstruct known HORs in human, it may report
incidental HORs in species, such as mouse and \emph{A. thaliana}, that only
have weak high-order patterns. We need to run TRF~\citep{Benson:1999aa} on SRF
contigs to obtain minimal repeat units. SRF may also assemble the same class of
repeat into multiple similar but not identical copies. We can align assembled
repeat units to identify such redundancy. Manual curation is recommended for a
deeper insight into the SatDNA structure of a new species.

Estimating the abundance of SatDNA is challenging. Sometimes ancient SatDNA
repeats may be too diverged from the assembled repeat consensus to be aligned
confidently. In human, whether to count scattered monomers in pericentromeric
regions as long SatDNA arrays would affect the estimate as well. In addition,
occasionally SatDNA units can be $>$5 kb in length. We may not observe clear
tandem patterns in $\sim$10 kb HiFi reads, which would lead to underestimate.
We do not have an automated algorithm to provide accurate abundance estimate in
corner cases.

SatDNA is pervasive in many species. It is however often underrepresented in
current reference genomes such as the human GRCh38 genome and the
\emph{Drosophila} BDGP6 genome. Even with improved sequencing technologies and
assembly algorithms, the assembly of SatDNA is often fragmented. With thousands
of species sequenced recently~\citep{Challis:2020aa,Rhie:2021ug} and more to
come in future, SRF may become an important tool to identity and annotate
SatDNA in these species. It may also supplement
RepeatModeler~\citep{Flynn:2020aa} to provide a more comprehensive repeat
library for masking SatDNA in assembled genomes.

\section{Methods}

\subsection{Running SRF for human assemblies}

We counted 171-mers occurring 20 times or more with KMC, using command line
{\tt kmc -fm -k171 -ci20 -cs100000} and extracted the 171-mer counts with
{\tt kmc\_dump}. SRF is directly applied to the output of {\tt kmc\_dump}
output in the default setting.

\subsection{Running SRF on sequence reads}

We estimated the approximate read depth by dividing the total number of read
bases by the number of bases in the reference genome or the corresponding
read assembly. We counted 151-mers with {\tt kmc -fq -k151 -ciXX -cs1000000},
where {\tt XX} is 10 times the average read depth of each sample.

\subsection{Estimating the abundance of SatDNA}

We aligned reconstructed repeat units to HiFi reads or contigs with
minimap2~\citep{Li:2018ab}, using command line {\tt minimap2 -c -N1000000
-f1000 -r100,100 <(srfutils.js enlong srf.fa)}, where {\tt srfutils.js} is a
companion script along with the SRF tool. Option {\tt -N1000000} asks minimap2
to report up to a million hits per query sequence; {\tt -f1000} considers
high-occurrence seeds; {\tt -r100,100} enables a small bandwidth of 100 bp
during alignment.

After the alignment, we used {\tt srfutils.js paf2bed} to filter poor
alignments and to merge adjacent alignments, and then used {\tt srfutils.js
bed2abun} to calculate the abundance of each repeat unit.

\subsection{Running TAREAN}

For human CHM13, we used real reads short reads reads. We ran TAREAN with {\tt
singularity exec --bind \$\{PWD\}:/data/ shub://repeatexplorer/repex\_tarean
seqclust -p -c 32 -r 50000000}. For \emph{Drosorphila}, we simulated 125 bp
paired-end reads from HiFi reads with {\tt dwgsim -N 146000 -1 125 -2 125 -y0
-e0 -E0 -r0 -F0 -R0}. This command line did not add additional sequencing
errors; the simulated reads only carried real sequencing errors on the original
HiFi reads.

\section{Data access}

The SRF implementation and associated analysis scripts are provided at
\href{https://github.com/lh3/srf}{https://github.com/lh3/srf}. A modified TRF
with an alternative command-line interface is available at
\href{https://github.com/lh3/TRF-mod}{https://github.com/lh3/TRF-mod}.
Assembled repeat units and their abundance estimates can be found at
\href{https://zenodo.org/record/7814465}{https://zenodo.org/record/7814465}.

\section{Competing interest statement}

H.L. is a consualtant for Integrated DNA Technologies, Inc.

\section{Acknowledgements}

This work is supported by US National Human Genome Research Institute (NHGRI)
grant R01HG010040 and U01HG010961 to H.L.

\bibliography{srf}

\begin{thebibliography}{}

\bibitem[Altemose, 2022]{Altemose:2022vw}
Altemose, N. (2022).
\newblock A classical revival: Human satellite dnas enter the genomics era.
\newblock {\em Semin Cell Dev Biol}, 128:2--14.

\bibitem[Altemose et~al., 2022]{Altemose:2022tv}
Altemose, N., Logsdon, G.~A., Bzikadze, A.~V., Sidhwani, P., Langley, S.~A.,
  Caldas, G.~V., Hoyt, S.~J., Uralsky, L., Ryabov, F.~D., Shew, C.~J., et~al.
  (2022).
\newblock Complete genomic and epigenetic maps of human centromeres.
\newblock {\em Science}, 376:eabl4178.

\bibitem[Ananiev et~al., 1998a]{Ananiev:1998aa}
Ananiev, E.~V., Phillips, R.~L., and Rines, H.~W. (1998a).
\newblock Chromosome-specific molecular organization of maize (zea mays l.)
  centromeric regions.
\newblock {\em Proc Natl Acad Sci U S A}, 95:13073--8.

\bibitem[Ananiev et~al., 1998b]{Ananiev:1998ab}
Ananiev, E.~V., Phillips, R.~L., and Rines, H.~W. (1998b).
\newblock Complex structure of knob dna on maize chromosome 9. retrotransposon
  invasion into heterochromatin.
\newblock {\em Genetics}, 149(4):2025--37.

\bibitem[Arora et~al., 2021]{Arora:2021aa}
Arora, U.~P., Charlebois, C., Lawal, R.~A., and Dumont, B.~L. (2021).
\newblock Population and subspecies diversity at mouse centromere satellites.
\newblock {\em BMC Genomics}, 22:279.

\bibitem[Benson, 1999]{Benson:1999aa}
Benson, G. (1999).
\newblock Tandem repeats finder: a program to analyze {DNA} sequences.
\newblock {\em Nucleic Acids Res}, 27:573--80.

\bibitem[Challis et~al., 2020]{Challis:2020aa}
Challis, R., Richards, E., Rajan, J., Cochrane, G., and Blaxter, M. (2020).
\newblock {BlobToolKit} - interactive quality assessment of genome assemblies.
\newblock {\em G3 (Bethesda)}, 10(4):1361--1374.

\bibitem[Cheng et~al., 2021]{Cheng:2021aa}
Cheng, H., Concepcion, G.~T., Feng, X., Zhang, H., and Li, H. (2021).
\newblock Haplotype-resolved de novo assembly using phased assembly graphs with
  hifiasm.
\newblock {\em Nat Methods}, 18:170--175.

\bibitem[Flynn et~al., 2020]{Flynn:2020aa}
Flynn, J.~M., Hubley, R., Goubert, C., Rosen, J., Clark, A.~G., Feschotte, C.,
  and Smit, A.~F. (2020).
\newblock Repeatmodeler2 for automated genomic discovery of transposable
  element families.
\newblock {\em Proc Natl Acad Sci U S A}, 117(17):9451--9457.

\bibitem[Foote et~al., 2022]{foote_genome_2022}
Foote, A., Bunskoek, P., {Wellcome Sanger Institute Tree of Life programme},
  {Wellcome Sanger Institute Scientific Operations: DNA Pipelines collective},
  {Tree of Life Core Informatics collective}, and {Darwin Tree of Life
  Consortium} (2022).
\newblock The genome sequence of the killer whale, {Orcinus} orca ({Linnaeus},
  1758).
\newblock {\em Wellcome Open Research}, 7:250.

\bibitem[Gao et~al., 2022]{Gao2022.08.07.502881}
Gao, S., Yang, X., Zhao, X., Wang, B., and Ye, K. (2022).
\newblock {HiCAT}: A tool for automatic annotation of centromere structure.
\newblock {\em bioRxiv}.

\bibitem[Hon et~al., 2020]{Hon:2020aa}
Hon, T., Mars, K., Young, G., Tsai, Y.-C., Karalius, J.~W., Landolin, J.~M.,
  Maurer, N., Kudrna, D., Hardigan, M.~A., Steiner, C.~C., Knapp, S.~J., Ware,
  D., Shapiro, B., Peluso, P., and Rank, D.~R. (2020).
\newblock Highly accurate long-read {HiFi} sequencing data for five complex
  genomes.
\newblock {\em Sci Data}, 7:399.

\bibitem[Khost et~al., 2017]{Khost:2017aa}
Khost, D.~E., Eickbush, D.~G., and Larracuente, A.~M. (2017).
\newblock Single-molecule sequencing resolves the detailed structure of complex
  satellite dna loci in drosophila melanogaster.
\newblock {\em Genome Res}, 27:709--721.

\bibitem[Kokot et~al., 2017]{Kokot:2017aa}
Kokot, M., Dlugosz, M., and Deorowicz, S. (2017).
\newblock {KMC} 3: counting and manipulating k-mer statistics.
\newblock {\em Bioinformatics}, 33(17):2759--2761.

\bibitem[Kunyavskaya et~al., 2022]{Kunyavskaya:2022tx}
Kunyavskaya, O., Dvorkina, T., Bzikadze, A.~V., Alexandrov, I.~A., and Pevzner,
  P.~A. (2022).
\newblock Automated annotation of human centromeres with {HORmon}.
\newblock {\em Genome Res}, 32:1137--1151.

\bibitem[Könyves et~al., 2022]{konyves_genome_2022}
Könyves, K., Mian, S., Johns, J., {Royal Botanic Garden Edinburgh Genome
  Acquisition Lab}, {Royal Botanic Gardens Kew Genome Acquisition Lab}, {Darwin
  Tree of Life Barcoding collective}, {Wellcome Sanger Institute Tree of Life
  programme}, {Wellcome Sanger Institute Scientific Operations: DNA Pipelines
  collective}, {Tree of Life Core Informatics collective}, Ruhsam, M., Leitch,
  I.~J., and {Darwin Tree of Life Consortium} (2022).
\newblock The genome sequence of the apple, {Malus} domestica ({Suckow})
  {Borkh}., 1803.
\newblock {\em Wellcome Open Research}, 7:297.

\bibitem[Lawniczak et~al., 2021]{lawniczak_genome_2021}
Lawniczak, M.~K., {Darwin Tree of Life Barcoding collective}, {Wellcome Sanger
  Institute Tree of Life programme}, {Wellcome Sanger Institute Scientific
  Operations: DNA Pipelines collective}, {Tree of Life Core Informatics
  collective}, and {Darwin Tree of Life Consortium} (2021).
\newblock The genome sequence of the spiny starfish, {Marthasterias} glacialis
  ({Linnaeus}, 1758).
\newblock {\em Wellcome Open Research}, 6:295.

\bibitem[Lawniczak et~al., 2022]{lawniczak_genome_2022}
Lawniczak, M.~K., {Darwin Tree of Life Barcoding collective}, {Wellcome Sanger
  Institute Tree of Life programme}, {Wellcome Sanger Institute Scientific
  Operations: DNA Pipelines collective}, {Tree of Life Core Informatics
  collective}, and {Darwin Tree of Life Consortium} (2022).
\newblock The genome sequence of the blue-rayed limpet, {Patella} pellucida
  {Linnaeus}, 1758.
\newblock {\em Wellcome Open Research}, 7:126.

\bibitem[Li, 2018]{Li:2018ab}
Li, H. (2018).
\newblock Minimap2: pairwise alignment for nucleotide sequences.
\newblock {\em Bioinformatics}, 34(18):3094--3100.

\bibitem[Li, 2019]{Li:2019aa}
Li, H. (2019).
\newblock Identifying centromeric satellites with dna-brnn.
\newblock {\em Bioinformatics}, 35:4408--4410.

\bibitem[Logsdon et~al., 2021]{Logsdon:2021aa}
Logsdon, G.~A., Vollger, M.~R., Hsieh, P., Mao, Y., Liskovykh, M.~A., Koren,
  S., Nurk, S., Mercuri, L., Dishuck, P.~C., Rhie, A., de~Lima, L.~G.,
  Dvorkina, T., Porubsky, D., Harvey, W.~T., Mikheenko, A., Bzikadze, A.~V.,
  Kremitzki, M., Graves-Lindsay, T.~A., Jain, C., Hoekzema, K., Murali, S.~C.,
  Munson, K.~M., Baker, C., Sorensen, M., Lewis, A.~M., Surti, U., Gerton,
  J.~L., Larionov, V., Ventura, M., Miga, K.~H., Phillippy, A.~M., and Eichler,
  E.~E. (2021).
\newblock The structure, function and evolution of a complete human chromosome
  8.
\newblock {\em Nature}, 593:101--107.

\bibitem[Lohse et~al., 2021]{lohse_genome_2021}
Lohse, K., García-Berro, A., Talavera, G., {Darwin Tree of Life Barcoding
  collective}, {Wellcome Sanger Institute Tree of Life programme}, {Wellcome
  Sanger Institute Scientific Operations: DNA Pipelines collective}, {Tree of
  Life Core Informatics collective}, and {Darwin Tree of Life Consortium}
  (2021).
\newblock The genome sequence of the red admiral, {Vanessa} atalanta
  ({Linnaeus}, 1758).
\newblock {\em Wellcome Open Research}, 6:356.

\bibitem[Lower et~al., 2018]{Lower:2018aa}
Lower, S.~S., McGurk, M.~P., Clark, A.~G., and Barbash, D.~A. (2018).
\newblock Satellite dna evolution: old ideas, new approaches.
\newblock {\em Curr Opin Genet Dev}, 49:70--78.

\bibitem[Melters et~al., 2013]{Melters:2013va}
Melters, D.~P., Bradnam, K.~R., Young, H.~A., Telis, N., May, M.~R., Ruby,
  J.~G., Sebra, R., Peluso, P., Eid, J., Rank, D., et~al. (2013).
\newblock Comparative analysis of tandem repeats from hundreds of species
  reveals unique insights into centromere evolution.
\newblock {\em Genome Biol}, 14:R10.

\bibitem[Miga, 2019]{Miga:2019aa}
Miga, K.~H. (2019).
\newblock Centromeric satellite dnas: Hidden sequence variation in the human
  population.
\newblock {\em Genes (Basel)}, 10.

\bibitem[Miga et~al., 2014]{Miga:2014aa}
Miga, K.~H., Newton, Y., Jain, M., Altemose, N., Willard, H.~F., and Kent,
  W.~J. (2014).
\newblock Centromere reference models for human chromosomes x and y satellite
  arrays.
\newblock {\em Genome Res}, 24:697--707.

\bibitem[Naish et~al., 2021]{Naish:2021aa}
Naish, M., Alonge, M., Wlodzimierz, P., Tock, A.~J., Abramson, B.~W.,
  Schm{\"u}cker, A., Mand{\'a}kov{\'a}, T., Jamge, B., Lambing, C., Kuo, P.,
  Yelina, N., Hartwick, N., Colt, K., Smith, L.~M., Ton, J., Kakutani, T.,
  Martienssen, R.~A., Schneeberger, K., Lysak, M.~A., Berger, F., Bousios, A.,
  Michael, T.~P., Schatz, M.~C., and Henderson, I.~R. (2021).
\newblock The genetic and epigenetic landscape of the arabidopsis centromeres.
\newblock {\em Science}, 374:eabi7489.

\bibitem[Newman et~al., 2022]{newman_genome_2022}
Newman, C., Tsai, M.-s., Buesching, C.~D., Holland, P. W.~H., Macdonald, D.~W.,
  {Darwin Tree of Life Consortium}, {University of Oxford and Wytham Woods
  Genome Acquisition Lab}, {Wellcome Sanger Institute Tree of Life programme},
  {Wellcome Sanger Institute Scientific Operations: DNA Pipelines collective},
  and {Tree of Life Core Informatics collective} (2022).
\newblock The genome sequence of the {European} badger, {Meles} meles
  ({Linnaeus}, 1758).
\newblock {\em Wellcome Open Research}, 7:239.

\bibitem[Nov{\'a}k et~al., 2017]{Novak:2017wx}
Nov{\'a}k, P., {\'A}vila~Robledillo, L., Kobl{\'\i}{\v z}kov{\'a}, A.,
  Vrbov{\'a}, I., Neumann, P., and Macas, J. (2017).
\newblock {TAREAN}: a computational tool for identification and
  characterization of satellite dna from unassembled short reads.
\newblock {\em Nucleic Acids Res}, 45:e111.

\bibitem[Nurk et~al., 2022]{Nurk:2022up}
Nurk, S., Koren, S., Rhie, A., Rautiainen, M., Bzikadze, A.~V., Mikheenko, A.,
  Vollger, M.~R., Altemose, N., Uralsky, L., Gershman, A., et~al. (2022).
\newblock The complete sequence of a human genome.
\newblock {\em Science}, 376:44--53.

\bibitem[Pemberton et~al., 2021]{pemberton_genome_2021}
Pemberton, J., Johnston, S.~E., Fletcher, T.~J., {Darwin Tree of Life Barcoding
  collective}, {Wellcome Sanger Institute Tree of Life programme}, {Wellcome
  Sanger Institute Scientific Operations: DNA Pipelines collective}, {Tree of
  Life Core Informatics collective}, and {Darwin Tree of Life Consortium}
  (2021).
\newblock The genome sequence of the red deer, {Cervus} elaphus {Linnaeus}
  1758.
\newblock {\em Wellcome Open Research}, 6:336.

\bibitem[Pertile et~al., 2009]{Pertile:2009aa}
Pertile, M.~D., Graham, A.~N., Choo, K. H.~A., and Kalitsis, P. (2009).
\newblock Rapid evolution of mouse y centromere repeat dna belies recent
  sequence stability.
\newblock {\em Genome Res}, 19:2202--13.

\bibitem[Rabanal et~al., 2022]{Rabanal:2022aa}
Rabanal, F.~A., Gr{\"a}ff, M., Lanz, C., Fritschi, K., Llaca, V., Lang, M.,
  Carbonell-Bejerano, P., Henderson, I., and Weigel, D. (2022).
\newblock Pushing the limits of hifi assemblies reveals centromere diversity
  between two arabidopsis thaliana genomes.
\newblock {\em Nucleic Acids Res}, 50:12309--12327.

\bibitem[Rautiainen et~al., 2023]{Rautiainen:2023aa}
Rautiainen, M., Nurk, S., Walenz, B.~P., Logsdon, G.~A., Porubsky, D., Rhie,
  A., Eichler, E.~E., Phillippy, A.~M., and Koren, S. (2023).
\newblock Telomere-to-telomere assembly of diploid chromosomes with verkko.
\newblock {\em Nat Biotechnol}.

\bibitem[Rhie et~al., 2021]{Rhie:2021ug}
Rhie, A., McCarthy, S.~A., Fedrigo, O., Damas, J., Formenti, G., Koren, S.,
  Uliano-Silva, M., Chow, W., Fungtammasan, A., Kim, J., Lee, C., Ko, B.~J.,
  Chaisson, M., Gedman, G.~L., Cantin, L.~J., Thibaud-Nissen, F., Haggerty, L.,
  Bista, I., Smith, M., Haase, B., Mountcastle, J., Winkler, S., Paez, S.,
  Howard, J., Vernes, S.~C., Lama, T.~M., Grutzner, F., Warren, W.~C.,
  Balakrishnan, C.~N., Burt, D., George, J.~M., Biegler, M.~T., Iorns, D.,
  Digby, A., Eason, D., Robertson, B., Edwards, T., Wilkinson, M., Turner, G.,
  Meyer, A., Kautt, A.~F., Franchini, P., Detrich, 3rd, H.~W., Svardal, H.,
  Wagner, M., Naylor, G. J.~P., Pippel, M., Malinsky, M., Mooney, M.,
  Simbirsky, M., Hannigan, B.~T., Pesout, T., Houck, M., Misuraca, A., Kingan,
  S.~B., Hall, R., Kronenberg, Z., Sovi{\'c}, I., Dunn, C., Ning, Z., Hastie,
  A., Lee, J., Selvaraj, S., Green, R.~E., Putnam, N.~H., Gut, I., Ghurye, J.,
  Garrison, E., Sims, Y., Collins, J., Pelan, S., Torrance, J., Tracey, A.,
  Wood, J., Dagnew, R.~E., Guan, D., London, S.~E., Clayton, D.~F., Mello,
  C.~V., Friedrich, S.~R., Lovell, P.~V., Osipova, E., Al-Ajli, F.~O.,
  Secomandi, S., Kim, H., Theofanopoulou, C., Hiller, M., Zhou, Y., Harris,
  R.~S., Makova, K.~D., Medvedev, P., Hoffman, J., Masterson, P., Clark, K.,
  Martin, F., Howe, K., Flicek, P., Walenz, B.~P., Kwak, W., Clawson, H.,
  Diekhans, M., Nassar, L., Paten, B., Kraus, R. H.~S., Crawford, A.~J.,
  Gilbert, M. T.~P., Zhang, G., Venkatesh, B., Murphy, R.~W., Koepfli, K.-P.,
  Shapiro, B., Johnson, W.~E., Di~Palma, F., Marques-Bonet, T., Teeling, E.~C.,
  Warnow, T., Graves, J.~M., Ryder, O.~A., Haussler, D., O'Brien, S.~J.,
  Korlach, J., Lewin, H.~A., Howe, K., Myers, E.~W., Durbin, R., Phillippy,
  A.~M., and Jarvis, E.~D. (2021).
\newblock Towards complete and error-free genome assemblies of all vertebrate
  species.
\newblock {\em Nature}, 592(7856):737--746.

\bibitem[Ruhsam et~al., 2022]{ruhsam_genome_2022}
Ruhsam, M., Bell, D., Hart, M., Hollingsworth, P., {Royal Botanic Garden
  Edinburgh Genome Acquisition Lab}, {Darwin Tree of Life Barcoding
  collective}, {Wellcome Sanger Institute Tree of Life programme}, {Wellcome
  Sanger Institute Scientific Operations: DNA Pipelines collective}, {Tree of
  Life Core Informatics collective}, and {Darwin Tree of Life Consortium}
  (2022).
\newblock The genome sequence of the {European} crab apple, {Malus} sylvestris
  ({L}.) {Mill}., 1768.
\newblock {\em Wellcome Open Research}, 7:296.

\bibitem[Schneider et~al., 2017]{Schneider:2017aa}
Schneider, V.~A., Graves-Lindsay, T., Howe, K., Bouk, N., Chen, H.-C., Kitts,
  P.~A., Murphy, T.~D., Pruitt, K.~D., Thibaud-Nissen, F., Albracht, D.,
  Fulton, R.~S., Kremitzki, M., Magrini, V., Markovic, C., McGrath, S.,
  Steinberg, K.~M., Auger, K., Chow, W., Collins, J., Harden, G., Hubbard, T.,
  Pelan, S., Simpson, J.~T., Threadgold, G., Torrance, J., Wood, J.~M., Clarke,
  L., Koren, S., Boitano, M., Peluso, P., Li, H., Chin, C.-S., Phillippy,
  A.~M., Durbin, R., Wilson, R.~K., Flicek, P., Eichler, E.~E., and Church,
  D.~M. (2017).
\newblock Evaluation of {GRCh38} and de novo haploid genome assemblies
  demonstrates the enduring quality of the reference assembly.
\newblock {\em Genome Res}, 27:849--864.

\bibitem[Sevim et~al., 2016]{Sevim:2016tl}
Sevim, V., Bashir, A., Chin, C.-S., and Miga, K.~H. (2016).
\newblock {Alpha-CENTAURI}: assessing novel centromeric repeat sequence
  variation with long read sequencing.
\newblock {\em Bioinformatics}, 32:1921--1924.

\bibitem[Shatskikh et~al., 2020]{Shatskikh:2020aa}
Shatskikh, A.~S., Kotov, A.~A., Adashev, V.~E., Bazylev, S.~S., and Olenina,
  L.~V. (2020).
\newblock Functional significance of satellite dnas: Insights from drosophila.
\newblock {\em Front Cell Dev Biol}, 8:312.

\bibitem[Sinding et~al., 2021]{sinding_genome_2021}
Sinding, M.-H.~S., Gopalakrishnan, S., Raundrup, K., Dalén, L., Threlfall, J.,
  {Darwin Tree of Life Barcoding collective}, {Wellcome Sanger Institute Tree
  of Life programme}, {Wellcome Sanger Institute Scientific Operations: DNA
  Pipelines collective}, {Tree of Life Core Informatics collective}, {Darwin
  Tree of Life Consortium}, and Gilbert, T. (2021).
\newblock The genome sequence of the grey wolf, {Canis} lupus {Linnaeus} 1758.
\newblock {\em Wellcome Open Research}, 6:310.

\bibitem[Thakur et~al., 2021]{Thakur:2021aa}
Thakur, J., Packiaraj, J., and Henikoff, S. (2021).
\newblock Sequence, chromatin and evolution of satellite dna.
\newblock {\em Int J Mol Sci}, 22.

\bibitem[Wei et~al., 2014]{Wei:2014vl}
Wei, K. H.-C., Grenier, J.~K., Barbash, D.~A., and Clark, A.~G. (2014).
\newblock Correlated variation and population differentiation in satellite
  {DNA} abundance among lines of drosophila melanogaster.
\newblock {\em Proc Natl Acad Sci U S A}, 111:18793--8.

\bibitem[Wood et~al., 2022]{wood_genome_2022}
Wood, C., Bishop, J., Harley, J., Mrowicki, R., {Marine Biological Association
  Genome Acquisition Lab}, {Darwin Tree of Life Barcoding collective},
  {Wellcome Sanger Institute Tree of Life programme}, {Wellcome Sanger
  Institute Scientific Operations: DNA Pipelines collective}, {Tree of Life
  Core Informatics collective}, and {Darwin Tree of Life Consortium} (2022).
\newblock The genome sequence of the orange-striped anemone, {Diadumene}
  lineata ({Verrill}, 1869).
\newblock {\em Wellcome Open Research}, 7:93.

\bibitem[Wright et~al., 2022]{wright_genome_2022}
Wright, R., Woof, K., Douglas, B., Gaya, E., {Royal Botanic Gardens Kew Genome
  Acquisition Lab}, {Darwin Tree of Life Barcoding collective}, {Wellcome
  Sanger Institute Tree of Life programme}, {Wellcome Sanger Institute
  Scientific Operations: DNA Pipelines collective}, {Tree of Life Core
  Informatics collective}, and {Darwin Tree of Life Consortium} (2022).
\newblock The genome sequence of the chicken of the woods fungus, {Laetiporus}
  sulphureus ({Bull}.) {Murrill}, 1920.
\newblock {\em Wellcome Open Research}, 7:83.

\bibitem[Zerbino and Birney, 2008]{Zerbino:2008uq}
Zerbino, D.~R. and Birney, E. (2008).
\newblock Velvet: algorithms for de novo short read assembly using de bruijn
  graphs.
\newblock {\em Genome Res}, 18:821--9.

\bibitem[Zhang et~al., 2019]{Zhang:2019ab}
Zhang, L., Hu, J., Han, X., Li, J., Gao, Y., Richards, C.~M., Zhang, C., Tian,
  Y., Liu, G., Gul, H., Wang, D., Tian, Y., Yang, C., Meng, M., Yuan, G., Kang,
  G., Wu, Y., Wang, K., Zhang, H., Wang, D., and Cong, P. (2019).
\newblock A high-quality apple genome assembly reveals the association of a
  retrotransposon and red fruit colour.
\newblock {\em Nat Commun}, 10(1):1494.

\end{thebibliography}

\end{document}